\documentclass[11pt]{article}
\usepackage{amsfonts}
\usepackage{amssymb}
\usepackage{graphicx}
\usepackage{amsmath}
\usepackage{color}
\usepackage[normalem]{ulem}

\newtheorem{theorem}{Theorem} [section]

\newtheorem{definition}[theorem]{Definition}
\newtheorem{example}[theorem]{Example}

\newtheorem{proposition}[theorem]{Proposition}

\newenvironment{proof}[1][Proof]{\textbf{#1.} }{\ \rule{0.5em}{0.5em}}
\allowdisplaybreaks[1]
\newcommand{\JP}[2]{{\color{blue}#1\ifmmode\msout{#2}\else\sout{#2}\fi}}
\newcommand{\AM}[2]{{\color{green}#1\ifmmode\msout{#2}\else\sout{#2}\fi}}
\newcommand{\LG}[2]{{\color{red}#1\ifmmode\msout{#2}\else\sout{#2}\fi}}
\begin{document}

\author{Luis A. Guardiola\thanks{%
Departamento de Fundamentos del An\'{a}lisis Econ\'{o}mico, Universidad de
Alicante, Alicante 03071, Spain. E-mail: luis.guardiola@ua.es}, Ana Meca%
\thanks{%
I.U. Centro de Investigaci\'{o}n Operativa. Universidad Miguel Hern\'{a}%
ndez, Edificio Torretamarit. Avda. de la Universidad s.n. 03202 Elche
(Alicante), Spain. E-mail: ana.meca@umh.es} and Justo Puerto\thanks{%
Facultad de Matem\'{a}ticas, Universidad de Sevilla, 41012 Sevilla, SPAIN.
E-mail: puerto@us.es}}
\title{Unitary Owen points in cooperative lot-sizing models with backlogging }
\maketitle

\begin{abstract}
Cooperative lot-sizing models with backlogging and heterogeneous costs are
studied in Guardiola et al. (2020). In this model several firms participate in a consortium aiming 
at satisfying their demand over the planing horizon with minimal operation cost. 
Each firm uses the best ordering channel  and holding technology provided by the 
participants in the consortium. The authors  show that there are always fair allocations
of the overall operation cost among the firms so that no group of agents profit from 
leaving the consortium. This paper revisits those cooperative lot-sizing models and 
presents a new family of cost allocations, the unitary Owen points. This family is an extension 
of the Owen set which enjoys very good properties in production-inventory proble, introduced by Guardiola et al. (2008).
Necessary and sufficient conditions are provided for the unitary Owen points to be fair 
allocations. In addition, we provide empirical evidence, throughout simulation, showing that the above condition is fulfilled in most cases.
Additionally, a relationship between lot-sizing games and a certain family 
of production-inventory games, through Owen's points of the latter, is described. This
interesting relationship enables to easily construct a variety of fair allocations for cooperative lot-sizing models. 
\bigskip

 \medskip

\end{abstract}

\newpage

\section{Introduction}

Lot-sizing problems with backlogging have been widely studied, during the
last 40 years, proving different reformulations that have made it possible
to find more efficiently solutions to the problem. Pochet and Wolsey (1988)
examine mixed integer programming reformulations of the uncapacitated
lot-sizing problem with backlogging which in an extended space of variables give strong reformulations using linear programming. 

More recently, a series of papers (Van Den Heuvel et al., 2007; Guardiola et al. 2008,
2009; Xu and Yang, 2009; Dreschel, 2010; Gopaladesikan and Uhan, 2011; Zeng
et al. 2011; Tsao et al. 2013; Chen and Zhang, 2016) have addressed the lot sizing problem under the perspective of cooperation considering cost sharing aspects of these models. Specifically, Van Den Heuvel
et al. (2007) focus on the cooperation in economic lot-sizing situations
with homogeneous costs but without backlogging (henceforth ELS-games).
Subsequently, Guardiola et al. (2008, 2009) present the class of
production-inventory games (henceforth, PI-games). Those papers provide a
cooperative approach to analyze the production and storage of indivisible
items being their characteristic function given as the optimal objective
function of a linear optimization problem. PI-games may be considered ELS games
without setup costs but with backlogging and heterogeneous costs. Guardiola
et al. (2009) prove that the Owen set, the set of allocations which are
achievable through dual solutions reduces to a singleton in the class of PI-games and these authors coined the term \textit{Owen point} to refer to that solution. The authors prove that the Owen point is always a
consistent core allocation. More recently, Chen and Zhang (2016) consider
the ELS-game with general concave ordering cost and they show that a core
allocation is computed in polynomial time under the assumption that all
retailers have the same cost parameters (again homogeneous costs). Their
approach is based on linear programming (LP) duality. Specifically, they
prove that there is an optimal dual solution that defines an allocation
in the core and point out that it is not necessarily true that every core allocation can be obtained by means of dual solutions.

Finally, setup-inventory games (henceforth, SI-games) are introduced in Guardiola
et. al. (2020) as a new class of combinatorial optimization games that
arises from cooperation in lot-sizing problems with backlogging and
heterogeneous costs. Each firm faces demand for a single product in each
period and coalitions can pool orders. Firms cooperate by using the best 
ordering channel and holding technology provided by the participants in the consortium, 
e.g. they produce, hold inventory, pay backlogged demand and make orders at the minimum cost among the members
of the coalition. Thus, firms aim at satisfying their demand over the
planing horizon with minimal operation cost. The authors show that there
are always fair allocations of the overall operation cost among the firms
so that no group of agents profit from leaving the consortium. That paper
proposes a parametric family of cost allocations and provides sufficient
conditions for this to be a stable family against coalitional defections of
firms. 

The Owen point works very well as long as one has a strong formulation for the underlying optimization problem, 
such as PI-problems, because the dual variables (shadow prices) are used to construct the core allocations.
However, this does not work for SI-problems, because in the original space of variables the corresponding optimization problem has integer
variables and strong duality does not apply. In this paper we extend further the idea of dual prices and we construct
an \textit{ad hoc}  kind of prices as the sum of the production, inventory and
backlogging costs plus a proportion of the fixed order cost which depends on
the total demand satisfied in that period. They are called unitary prices.
These unit prices enable to replicate the construction of the Owen point
by multiplying such unit prices by the demands and adding in all the
periods. These allocations ``\textit{a la Owen}" are called unitary Owen points.
Unfortunately one cannot always guarantee that unitary Owen points are core
allocations. Nevertheless, we provide necessary and sufficient conditions for this situation
to hold, i.e. for unitary Owen points to be core allocations and also show by simulation empirical evidence that this condition is satisfied in most cases.
Furthermore, we consider whether  it is possible to relate general SI-situations to simpler situations
where the core is well-known and characterized as in PI-games. In this regard, we prove 
that the answer to this question is yes: one can use the Owen point of the surplus game a PI-game which measures the excess in costs that occurs with respect to the
minimum unit price.

The contribution of this paper to the literature of lot-sizing games is twofold: firstly, a new family of cost allocations 
in cooperative lot-sizing models with backlogging and heterogeneous costs is presented: The unitary Owen points. 
It is an extension of the Owen set which enjoys very-good properties in production-inventory problems. Necessary and sufficient 
conditions are provided for unitary Owen points to be fair allocations. Furthermore, we empirically show that these conditions are satisfied for almost any SI-situation, resulting in an explicit quasi-solution for this class of games. Secondly, a relationship between lot-sizing games and a certain family of production-inventory  games, through Owen's points of the latter, is described. This interesting relationship enables us to analyze  cooperative lot-sizing models using properties of the much simpler and well-known class of PI-games.

The rest of this paper is organized as follows: the next section formulates SI-problems 
and shows that SI-games are totally balanced resorting to a result
of Pochet and Wolsey (1988). Section three describes the unitary Owen points, provides a 
necessary and sufficient condition for those points to be core allocations and gives empirical evidence to consider the unitary Owen point as a quasi-solution for SI-games. Section four presents a relationship between SI-games and a certain
family of PI-games through the Owen's points of the latter. This interesting relationship
simplifies the analysis and construction of core allocations for SI-games. 
Finally, the fifth section presents a research summary and some conclusions.

\section{SI-games: reformulation and balancedness}

We begin by formulating the setup-inventory problems with backlogging (SI-problems).

Consider $T$ periods, numbered from $1$ to $T$, where the demand
for a single product occurs in each of them. This demand is satisfied by own production, and can be done during the production periods, in previous periods (inventory) or later periods (backlogging). In each production period a fixed cost must be paid. Therefore, the model includes production, inventory holding, backlogging and setup costs. The aim is to find an optimal ordering plan, that is a feasible ordering plan that minimizes the sum of setup, production, inventory holding and backlogging cost. Although the model assumes that companies produce their demand, we can interchangeably consider the case where demand is satisfied either by producing or purchasing. One has simply to interpret that the purchasing costs can be ordering costs (set up costs) and unit purchasing costs (variable costs).
 The goal is to establish an operational plan in order to satisfy
demand at minimum total cost. Formally, for each period $t=1,\ldots ,T$ we
define $d_{t}$ as the integer demand to be satisfied, and $k_{t}$, $h_t$, $%
b_t$, $p_t$, respectively, as the setup, inventory carrying, backlogging and
unit production costs.

The decision variables of the model for each period $t$ are the order size $%
q_{t}$, the inventory at the end of the period $I_{t}$ and the backlogged
demand $E_{t}$. In addition, the order size must be integer. In the
following, we present a first mathematical programming formulation for the
setup-inventory problem (SI-problem). Let $M=\sum_{t=1}^{T}d_{t}$ and $z_{t}$
be a decision variable that assumes the value one if an order is placed
at the beginning of period $t$ and denote by $C(d,k,h,b,p)$ the minimum
overall  operation cost during the planning horizon, then 
\begin{eqnarray*}
C(d,k,h,b,p):=\quad &\min &\displaystyle\sum_{t=1}^{T}\left(
p_{t}q_{t}+h_{t}I_{t}+b_{t}E_{t}+k_{t}z_t\right)  \notag \\
&\mbox{s.t.}&I_{0}=I_{T}=E_{0}=E_{T}=0, \\
&&I_{t}-E_{t}=I_{t-1}-E_{t-1}+q_{t}-d_{t},\quad t=1,\ldots ,T, \\
&& q_t\le M \cdot z_{t}, \quad t=1,\ldots ,T, \\
&&q_{t},\;I_{t},\;E_{t},\mbox{ non-negative, integer},\;t=1\ldots ,T, \\
&&z_{t} \in \{0,1\}
\end{eqnarray*}
The above objective function minimizes the sum of all the considered costs over the planning horizon. The first constraint imposes that the model must start and finish with empty inventory. The second group of constraints are flow conservation constraints ensuring the right transition of inventory and backlogged demand among periods. The third group of constraints model that setup cost is only charged whenever an order is placed.

In order to simplify the notation we define $Z$ as a matrix in which all
costs are included, that is, $Z:=(K,H,B,P)$ being $K,H,B$ and $P$ the
matrices containing the setup, inventory carrying, backlogging and unit
production costs for all periods $t=1,\ldots,T$. A cost TU-game is a pair $%
(N,c)$, where $N$ is the finite player set, $\mathcal{P}(N)$ is
the power set of N (i.e. the set of coalitions in N), and $c:\mathcal{P}(N)\rightarrow 
\mathbb{R}$ the characteristic function satisfying $c(\varnothing )=0.$ A
cost allocation will be $x\in \mathbb{R}^{n}$ and, for every coalition $%
S\subseteq N$ we denote by $x(S)$ the aggregate cost-sharing for coalition $%
S $, i.e., $x(S):=\sum_{i\in S}x_{i}$ with $x(\varnothing )=0.$

For each SI-situation represented by its cost matrices $(N,D,Z)$, we associate a cost TU-game $(N,c)$ where,
for any nonempty coalition $S\subseteq
N,c(S):=C(d^{S},k^{S},h^{S},b^{S},p^{S})$ with $d^{S}=\sum_{i\in S}d^{i}$, where $d^{i}=\left(d^{i}_{1},...,d^{i}_{T}\right)$,
and the rest of the costs will be the minimum value among all the costs of
the players in the coalition $S$ at each one of the periods, serve as an
example $p^{S}=[p_{1}^{S},\ldots ,p_{T}^{S}]^{\prime }$ where $p_{t}^{S}=\min_{i\in S}\{p_{t}^{i}\}$ for $t=1,\ldots ,T.$ Every cost
TU-game defined in this way is what we call a setup-inventory game
(SI-game). The reader may notice that every PI-game (as introduced by Guardiola et
al., 2009) is a SI-game with $k_{t}=0,$ for all $t \in T.$ Moreover, as mentioned above, although the model assumes that companies produce their demand, we can interchangeably consider the case where demand is satisfied either by producing or purchasing. One has simply to interpret that the purchasing costs can be ordering costs (set up costs) and unit purchasing costs (variable costs).

Recall that the core of a game $(N,c)$ consists of those cost allocations which divide
the cost of the grand coalition, $c(N)$,  in such a way that any other coalition pays
at most its cost by the characteristic function. Formally,
$$Core(N,c)=\left\{ x\in \mathbb{R}^{n}\left/ x(N)=c(N)\text{ and }x(S)\leq
c(S)\text{ for all }S\subset N\right. \right\}.$$ 

In the following, fair allocations of the total cost will be called
core allocations. Bondareva (1963) and Shapley (1967) independently provide
a general characterization of games with a non-empty core by means of
balancedness. They prove that $(N,c)$ has a nonempty core if and only if it
is balanced. In addition, it is a totally balanced game\footnote{%
Totally balanced games were introduced by Shapley and Shubik in the study of
market games (see Shapley and Shubik, 1969).} if the core of every subgame
is nonempty.

Our goal is to show that SI-games are totally balanced resorting.
To do so we use an easy proof which is based on
duality resorting to a result by Pochet and Wolsey (1988). Observe that 
the characteristic function $c(S)$ of these games can be written as the optimal value of the following 
$LSI(S)$ problem: 
 
 \begin{align}
c(S)= \min & \sum_{1\le t\le T} \sum_{1\le \tau \le T} d_\tau^S p_{t\tau}^S \lambda_{t\tau} + \sum_{1\le t \le T} k_t^S z_t \tag{$LSI(S)$} \label{primal_S}\\
 \mbox{s.t. } & d_\tau ^S \sum_{1\le t\le T} \lambda_{t\tau}  = d_\tau^S , \quad \forall \tau=1,\ldots,T,\nonumber\\
 & \lambda_{t\tau}  \le z_t, \quad \forall t,\tau=1,\ldots,T,\nonumber \\
 & \lambda_{t\tau} ,z_t \in \{0,1\}. \nonumber
 \end{align}
 The variables $\lambda_{t\tau}$ are equal to 1 if and only if demand in period $\tau$ is produced in period $t$ and zero otherwise. Likewise, the variables $z_t$ are equal to 1 if and only if there is some production at period $t$. The cost of covering the demand in period  $\tau$ if the production is done in period $t$ is given by
\begin{equation} p_{t\tau}^S= \left\{ \begin{array}{ll} p_t^S & \mbox{if } t=\tau,\\
 p_t^S+\sum_{i=t}^{\tau -1}h_i^S & \mbox{if } t<\tau ,\\
 p_t^S+\sum_{i=\tau}^{t -1} b_i^S & \mbox{if } t>\tau.
 \end{array}\right.  \label{eq:ptt}
 \end{equation}
 This is the facility location reformulation by Pochet and Wolsey (1988) of the SI problem. This formulation has a strong dual if the underlying graph of the location problem is a tree (Tamir 1992). In this case, the graph is a line and thus the mentioned result applies. Let $y_\tau$ be the dual variable associated with the first constraints and $\beta_{t\tau}$ those associated with the second family of constraints, then the dual is: 
\begin{align}
c(S)=\max & \sum_{1\le \tau \le T} d_\tau ^S y_\tau  \label{dual_S} \tag{$DSI(S)$}\\
\mbox{s.t. } & \sum_{1\le \tau \le T} d_\tau ^S\beta_{t\tau} \le k_t^S, \forall t,\nonumber\\
& y_\tau - \beta_{t\tau} \le p_{t\tau}^S,\quad \forall t,\tau,\nonumber\\
& \beta_{t,\tau}\ge 0 , \; y_t \mbox{ free}. \nonumber
\end{align} 

 Pochet and Wolsey (1988) proved that the linear relaxation of a SI-problem, $%
LSI(S),$ has an integral optimal solution. Hence, the optimal value of its dual
problem matches that of the primal one, that is, $%
v(DSI(S))=C(d^{S},k^{S},h^{S},b^{S},p^{S})=c(S)$ for all $S\subseteq N.$

\begin{theorem}
\label{Balanced} \bigskip Every SI-game is totally balanced.
\end{theorem}

\begin{proof}
Take a SI-situation $(N,D,Z)$ and the associated SI-game $(N,c)$. Consider $%
(y^{\ast },\beta^{\ast })$ an optimal solution to dual $DSI(N)$ where $y^{\ast
}=(y_{1}^{\ast },...,y_{T}^{\ast })$ and $\beta^{\ast }=(\beta_{11}^{\ast
},...,\beta_{TT}^{\ast }).$ It is known from optimality that  
\begin{equation*}
\sum_{t=1}^{T}y_{t}^{\ast }d_{t}^{N}=v(DSI(N))=c(N)
\end{equation*}%
Note that the solution $(y^{\ast },\beta^{\ast })$ is also feasible for any dual
problem with $S\subseteq N$ since $p^{N}\leq p^{S},$ $h^{N}\leq h^{S},$ $%
b_{t}^{N}\leq b_{t}^{S}$ and $k_{t}^{N}\leq k_{t}^{S}.$ Therefore, 
\begin{equation*}
\sum_{t=1}^{T}y_{t}^{\ast }d_{t}^{S}\leq v(DSI(S))=c(S)
\end{equation*}%
Thus, the allocation $(\sum_{t=1}^{T}y_{t}^{\ast }d_{t}^{i})_{i\in N}\in
Core(N,c).$\newline
Note that every subgame of a SI-game is also a SI-game. Hence, we can also conclude that every SI-game is totally balanced.
\end{proof}

\section{Unitary Owen points}

In this section we introduce a new family of cost allocations on the class
of SI-games. This family is inspired by the flavour of the Owen 
point and its relationship with the shadow prices of the dual problems associated with
SI-problems. To define those cost allocations, it is necessary to describe 
the set of optimal plans and the unitary prices.

Consider a SI-situation $(N,D,Z)$. A feasible ordering plan for such a
situation is defined by $\sigma \in \mathbb{R}^{T}$ where $\sigma _{t}\in
T\cup \{0\}$ denotes the period where demand of period $t$ is ordered. We
assume the convention that $\sigma _{t}=0$ if and only if $d_{t}=0$. It
means that there is no order placed to satisfy demand at period $t$ since
demand at this period is null. Moreover, $P^{S}(\sigma )\in \mathbb{R}^{T}$ is defined 
as the operation cost vector associated to the ordering plan $\sigma $
(henceforth: cost-plan vector) for any coalition $S\subseteq N$, where

\begin{equation*}
P_{t}^{S}(\sigma )=\left\{ 
\begin{array}{ccc}
0 & \text{if} & \sigma _{t}=0, \\ 
p_{\sigma _{t}t}^{S} & \text{if} & \sigma _{t}\in \{1,...,T\}.%
\end{array}%
\right. 
\end{equation*}

Given an optimal ordering plan, $\sigma ^{S}$, for the SI-problem $C(d^{S},k^{S},h^{S},b^{S},p^{S})$, the characteristic function
is rewritten as follows: for  any non-empty coalition $S\subseteq N$,  
\begin{equation*}
c(S)=P^{S}(\sigma ^{S})^{\prime }d^{S}+\delta (\sigma ^{S})^{\prime
}k^{S}=\sum_{t=1}^{T}\left( P_{t}^{S}(\sigma ^{S})d_{t}^{S}+\delta
_{t}(\sigma ^{S})k_{t}^{S}\right) ,
\end{equation*}%
\noindent where, $\delta (\sigma ^{S})=\left( \delta _{t}(\sigma
^{S})\right) _{t\in T}$ and 
\begin{equation*}
\delta _{t}(\sigma ^{S})=\left\{ 
\begin{array}{cc}
1 & \text{if }\exists r\in T/\sigma _{r}^{S}=t\text{ }, \\ 
0 & \text{otherwise.}%
\end{array}%
\right.
\end{equation*}

The set of optimal plans is denoted by $\Lambda (N,D,Z):=\left\{ \left(
\sigma ^{S}\right) _{S\in \mathcal{P}(N)}\right\} $ where $\sigma ^{S}$ is
an optimal ordering plan associated to $LSI(S)$. Note that the set of optimal plans may be large since often there are
multiple optimal solutions for the program $LSI(S)$.  Core allocations built from optimal dual variables are known to exhibit some questionable properties as pointed out for instance by Perea et al. 2012 or Fern\'andez et. al. 2002. For this reason, whenever the core is larger than the set of allocations coming from dual variables, it is interesting to provide some alternative core allocations. In the following we derive alternatives which under mild conditions are stable, i.e. core allocations for these situations.

We define the \textit{unitary prices }
as the sum of the production, inventory and backlogging
costs plus a proportion of the fixed order cost which depends on the total
demand satisfied in each period.

\begin{definition}
Let $(N,D,Z)$ be a SI-situation and $\left( \sigma ^{S}\right) _{S\in 
\mathcal{P}(N)}\in \Lambda (N,D,Z).$ For each period $t\in T $ and each
coalition $S\subseteq N,$ the unitary price is defined as follows: 
\begin{equation*}
y_{t}\left( \sigma ^{S},d^{S},z^{S}\right) :=\left\{ 
\begin{array}{ccc}
0 & \text{if} & \sigma _{t}^{S}=0, \\ 
P_{t}^{S}(\sigma ^{S})+\frac{k_{\sigma _{t}^{S}}^{S}}{\sum_{m\in
Q^{S}(\sigma _{t}^{S})}d_{m}^{S}} & \text{if} & \sigma _{t}^{S}\neq 0,%
\end{array}%
\right.
\end{equation*}

where $Q^{S}(t):=\left\{ k\in T:\sigma _{k}^{S}=t\right\} $ and $z^{S}$
represents the cost matrix $(k^{S},h^{S},b^{S},p^{S}).$
\end{definition}
\medskip

The reader should observe that $Q^{S}(t)$ is the set of periods that satisfy the demand in period $t$, for the optimal plan $\sigma ^{S}$. 
Note that for any coalition $S\subseteq N,$ $\sum_{t=1}^{T}y_{t}\left(
\sigma ^{S},d^{S},z^{S}\right) \cdot d_{t}^{S}=c(S).$

The next proposition shows that we may construct core allocations from 
the unitary prices of the grand coalition as long as they are the cheapest in
each period with positive demand. We shall call them \emph{unitary Owen points.}

\begin{definition}
Let $(N,D,Z)$ be a SI-situation and $\left( \sigma ^{S}\right) _{S\in 
\mathcal{P}(N)}\in \Lambda (N,D,Z).$ The unitary Owen point is given by 
\begin{equation*}
\theta \left( \sigma ^{N},d^{N},z^{N}\right) :=\left(
\sum_{t=1}^{T}y_{t}\left( \sigma ^{N},d^{N},z^{N}\right) \cdot
d_{t}^{i}\right) _{i\in N}.
\end{equation*}
\end{definition}

Note that every optimal plan generates a unit price for each period of time and hence, a unitary Owen point.

Observe that from  $y(\sigma^N,d^N,z^N)$ we can build a solution $(y(\sigma^N,d^N,z^N), \beta(\sigma^N))$ with $\beta_{\sigma^N(\tau),\tau}=0$ if $\sigma^N(\tau)=0$  and $\beta_{\sigma^N(\tau),\tau}= \frac{k_{\sigma^N(\tau)}^N}{\sum_{m\in Q^N(\sigma^N(\tau))} d_m^N}$ if $\sigma^N(\tau)\neq 0$ satisfying $c(N)=\sum_{\tau=1}^T d_\tau^N y_t(\sigma^N,d^N,z^N)$. However, it may not be a feasible solution of the dual for the grand coalition \label{dual_S} whenever $P_\tau^N(\sigma^N(\tau))>p_{t\tau} ^N$ for some $t$. 
Still, the unitary Owen point associated with this dual solution can be a core allocation.

The following example elaborates on a SI-situation with 3 players and 2 periods. The unitary Owen point for the corresponding SI-game is a core allocation but this allocation does not come from optimal dual prices.

\begin{example}
\bigskip Consider the following SI-situation with two periods and
three players and the associated SI-game: 
\begin{equation*}
\begin{array}{|c|c|c||c|c||c||c||c|c||c|}
\hline
& d_{1}^{_{S}} & d_{2}^{_{S}} & p_{1}^{_{S}} & p_{2}^{_{S}} & h_{1}^{_{S}} & 
b_{1}^{_{S}} & k_{1}^{_{S}} & k_{2}^{_{S}} & c \\ \hline
\{1\} & 2 & 1 & 9 & 9 & 6 & 4 & 6 & 8 & 39 \\ \hline
\{2\} & 8 & 2 & 9 & 6 & 9 & 7 & 7 & 9 & 100 \\ \hline
\{3\} & 6 & 1 & 5 & 6 & 3 & 5 & 6 & 10 & 44 \\ \hline
\{1,2\} & 10 & 3 & 9 & 6 & 6 & 4 & 6 & 8 & 122 \\ \hline
\{1,3\} & 8 & 2 & 5 & 6 & 3 & 4 & 6 & 8 & 62 \\ \hline
\{2,3\} & 14 & 3 & 5 & 6 & 3 & 5 & 6 & 9 & 100 \\ \hline
\{1,2,3\} & 16 & 4 & 5 & 6 & 3 & 4 & 6 & 8 & 118 \\ \hline
\end{array}%
\end{equation*}

The optimal plan for coalition $N$ is $\sigma ^{N}=(1,1)$ with $p^{N}(\sigma
^{N})=\left( 5,8\right) $ and $y\left( \sigma ^{N},d^{N},z^{N}\right)
=\left( 5+\frac{3}{10},8+\frac{3}{10}\right) $. The unitary Owen point $%
\theta \left( {\sigma }^{N},d^{N},z^{N}\right) =\left( 18+\frac{9}{10},59,40+%
\frac{1}{10}\right) \in Core(N,c).$ Note that {$(y(\sigma
^{N},d^{N},z^{N}),\beta (\sigma ^{N}))$ with} $\beta _{21}({\sigma }^{N})=%
\frac{3}{10},\beta _{22}({\sigma }^{N})=\frac{3}{10}$ and  $\beta
_{t\tau }({\sigma }^{N})=0$ for all the remaining $t$ and $\tau$,  is not feasible for the dual problem  $DSI(N)$. Indeed, it violates the constraint $y_{2}({\sigma }^{N},d^{N},z^{N})-\beta _{22}(\sigma ^{N})\leq p_{22}^{N}$,
since this is equivalent to $p_{2}^{N}(\sigma ^{N})\leq p_{22}^{N}$
but $p_{2}^{N}(\sigma ^{N})=8$ and $p_{22}^{N}=6$.    
\end{example}

Therefore, it is clear that the unitary Owen point can provide core allocation which do not come from optimal dual prices, although it is not clear under which conditions this unitary price fulfills this property. The following result provides an easy sufficient condition for this to happen.

\begin{proposition}
\label{prop1} Let $(N,D,Z)$ be a SI-situation, $\left( \sigma ^{S}\right)
_{S\in \mathcal{P}(N)}\in \Lambda (N,D,Z)$, and $(N,c)$ the associated
SI-game. If $\ y_{t}\left( \sigma ^{N},d^{N},z^{N}\right) \leq y_{t}\left(
\sigma ^{S},d^{S},z^{S}\right) $ for all $t\in T$ and for all $S\subset N$
with $d_{t}^{S}\neq 0$, then 
\begin{equation*}
\theta \left( \sigma ^{N},d^{N},z^{N}\right) \in Core(N,c).
\end{equation*}
\end{proposition}

\begin{proof}
It is straightforward from the definition of the unitary Owen point.
\end{proof}
\medskip

It would be reasonable that the larger a coalition the lower its unit prices, since its members operate with the best technology available in
the group. Unfortunately, this condition is not always satisfied 
as Example \ref{example-2P} shows. Therefore, we are interested in 
finding stronger conditions than the one given in Proposition \ref{prop1}. In the following we address this question.

\medskip

In order to simplify the notation, for each $t\in T,$ we define:

\begin{itemize}
\item Cost difference per demand unit between coalition $S$ and $R$ in a
period $t$: 
\begin{equation*}
a_{t}^{SR}:=P_{t}^{S}(\sigma ^{S})-P_{t}^{R}(\sigma ^{R}).
\end{equation*}
Note that $a_{t}^{SR}+a_{t}^{RS}=0.$

\item Aggregate demand of coalition $S \subseteq N$ in all those periods
that satisfy its demand in period $t$: 
\begin{equation*}
\alpha _{t}(S):=\sum_{m\in Q^{N}(t)}d_{m}^{S}.
\end{equation*}

\item Aggregate order cost of coalition $S \subseteq N:$ 
\begin{equation*}
k(S):=\sum_{t\in T^{S}}k_{t}^{S},
\end{equation*}
where $T^{S}:=\left\{ t\in T\: \delta _{t}(\sigma ^{S})=1
\right\} $ is the set of ordering periods.
\end{itemize}

The next theorem provides necessary and sufficient conditions for the unitary Owen point to be a core
allocation. These conditions state an upper bound for the average cost savings per unit demand in the grand coalition, 
for those periods where an order is placed. Such an upper bound is related to the savings in fixed order costs. 

\begin{theorem}
\label{prop2 copy(1)} Let $(N,D,Z)$ be a SI-situation, $\left( \sigma
^{S}\right) _{S\in \mathcal{P}(N)}\in \Lambda (N,D,Z)$, and $(N,c)$ the
associated SI-game. $\theta \left( \sigma ^{N},d^{N},z^{N}\right) \in Core(N,c)$
if and only if there are real weights $\beta_{t}^{S}$, for any {$S\varsubsetneq N$} and 
any $t\in T^{N}$ with $\alpha _{t}(S)>0$,  satisfying that 

\begin{equation*}
\sum_{j\in Q^{N}(t)}\frac{a_{j}^{NS}\cdot d_{j}^{S}}{\alpha _{t}(S)}\leq
\beta _{t}^{S}\cdot \frac{k(S)}{\alpha _{t}(S)}-\frac{k_{t}^{N}}{\alpha
_{t}(N)}
\end{equation*}%
with $\sum_{t\in T^{N}:\alpha _{t}(S)>0}\beta _{t}^{S}\leq 1.$
\end{theorem}

\begin{proof}
\textbf{(if)} Take $\left( \sigma ^{S}\right) _{S\in \mathcal{P}(N)}\in \Lambda
(N,D,Z)$ and consider a coalition {$S\varsubsetneq N.$ We must prove that 
$\theta \left( \sigma ^{N},d^{N},z^{N}\right) \in Core(N,c)$, e.g.  
$\sum_{i\in S}\theta _{i}\left( \sigma ^{N},d^{N},z^{N}\right) -c(S)\leq 0.$
Indeed,  
$$\sum_{i\in S}\theta _{i}\left( \sigma ^{N},d^{N},z^{N}\right) -c(S)$$

$$=\sum_{t=1}^{T}y_{t}\left( \sigma ^{N},d^{N},z^{N}\right) \cdot
d_{t}^{S}-\sum_{t=1}^{T}y_{t}\left( \sigma ^{S},d^{S},z^{S}\right) \cdot
d_{t}^{S}$$
$$=\sum_{t=1}^{T}\left( a_{t}^{NS}\cdot d_{t}^{S}+\frac{k_{\sigma
_{t}^{N}}^{N}\cdot d_{j}^{S}}{\sum_{m\in Q^{N}(\sigma _{t}^{N})}d_{m}^{N}}-%
\frac{k_{\sigma _{t}^{S}}^{S}\cdot d_{j}^{S}}{\sum_{m\in Q^{S}(\sigma
_{t}^{S})}d_{m}^{S}}\right)$$
$$=\sum_{t\in T^{N}}\left( \sum_{j\in Q^{N}(t)}\left( a_{j}^{NS}\cdot
d_{j}^{S}+\frac{k_{t}^{N}\cdot d_{j}^{S}}{\alpha _{t}(N)}\right) \right)
-k(S)$$
$$=\sum_{t\in T^{N}:\alpha _{t}(S)>0}\left( \alpha _{t}(S)\cdot \sum_{j\in
Q^{N}(t)}\left( \frac{a_{j}^{NS}\cdot d_{j}^{S}}{\alpha _{t}(S)}\right) +%
\frac{k_{t}^{N}\cdot \alpha _{t}(S)}{\alpha _{t}(N)}\right) -k(S)$$
$$\leq \sum_{t\in T^{N}:\alpha _{t}(S)>0}\left( \beta _{t}^{S}\cdot \frac{%
\alpha _{t}(S)\cdot k(S)}{\alpha _{t}(S)}-\frac{\alpha _{t}(S)\cdot k_{t}^{N}%
}{\alpha _{t}(N)}+\frac{k_{t}^{N}\cdot \alpha _{t}(S)}{\alpha _{t}(N)}%
\right) -k(S)$$
$$=k(S)\cdot \sum_{t\in T^{N}:\alpha _{t}(S)>0}\beta _{t}^{S}-k(S)\leq 0$$


\textbf{(only if)} Consider now that $\theta \left( \sigma ^{N},d^{N},z^{N}\right) \in
Core(N,c).$ Then, for all }$S\subset N,$ $\sum_{i\in S}\theta _{i}\left( \sigma
^{N},d^{N},z^{N}\right) -c(S)\leq 0$ which  is
equivalent to
\begin{equation*}
\sum_{t\in T^{N}}\left( \sum_{j\in Q^{N}(t)}\left( a_{j}^{NS}\cdot d_{j}^{S}+%
\frac{k_{t}^{N}\cdot d_{j}^{S}}{\alpha _{t}(N)}\right) \right) \leq k(S).
\end{equation*}

For all $t\in T^{N}$ and every coalition $S\varsubsetneq N$ 
with $\alpha _{t}(S)>0$ there are always real weights $\beta _{t}^{S}$ with 
$\sum_{t\in T^{N}}\beta _{t}^{S}\leq 1,$ satisfying  
\begin{align*}
\sum_{j\in Q^{N}(t)}\left( a_{j}^{NS}\cdot d_{j}^{S}+\frac{k_{t}^{N}\cdot
d_{j}^{S}}{\alpha _{t}(N)}\right)&  \leq \beta _{t}^{S}\cdot k(S),\\
\sum_{j\in Q^{N}(t)}\left( \frac{a_{j}^{NS}\cdot d_{j}^{S}}{\alpha _{t}(S)}%
\right) +\frac{k_{t}^{N}\cdot \alpha _{t}(S)}{\alpha _{t}(N)\cdot \alpha
_{t}(S)}  & \leq \frac{\beta _{t}^{S}\cdot k(S)}{\alpha _{t}(S)}, \\
\sum_{j\in Q^{N}(t)}\frac{a_{j}^{NS}\cdot d_{j}^{S}}{\alpha _{t}(S)} &
\leq \beta _{t}^{S}\cdot \frac{k(S)}{\alpha _{t}(S)}-\frac{k_{t}^{N}}{\alpha
_{t}(N)}.
\end{align*}
\end{proof}

At first glance, the reader might think that the conditions of the previous theorem are too restrictive, i.e. they are only satisfied by a small family of SI-situations.  However, an empirical analysis simulating SI-situations  shows that most of the instances satisfy those conditions. Indeed, we start by randomly generating (using the uniform probability distribution) a first set of $100,000$ instances of  SI-situations so that for every player and 
for each period the data range in $d_{t}^{i}\in\left[0,30\right]$, $p_{t}^{i},h_{t}^{i},b_{t}^{i}\in\left[0,10\right]$ and $k_{t}^{i}\in\left[0,50\right]$.
The percentage of SI-situations for which the Unitary Owen point belongs to the core of the corresponding SI-game is shown in Table \ref{t:tabla1}.
\renewcommand{\arraystretch}{1.5}
\begin{table}[htb]
\[
\begin{tabular}{|c|c|c|c|c|}
\hline
Players & $T=2$ & $T=3$ & $T=4$ & $T=5$ \\ \hline
2 & $99.934\%$ & $99.979\%$ & $99.993\%$ & $100\%$  \\ 
3 & $99.942\%$ & $99.983\%$ & $99.989\%$ & $99.995\%$  \\ 
4 & $99.950\%$ & $99.991\%$ & $99.996\%$ & $99.999\%$  \\ 
5 & $99.974\%$ & $99.982\%$ & $99.992\%$ & $99.998\%$ \\ 
6 & $99.974\%$ & $99.993\%$ & $99.998\%$ & $99.999\%$  \\ 
7 & $99.985\%$ & $99.996\%$ & $99.999\%$ & $100\%$  \\ \hline
\end{tabular}%
\]
\caption{Percentage of instances fulfilling the condition of Theorem \ref{prop2 copy(1)} for the first set of instances\label{t:tabla1}}
\end{table}

It can be seen that the the larger the number of players and periods the higher the percentage that some unitary Owen point belongs to the core. In case that we impose that the demand and the costs are greater than zero: $d_{t}^{i}\in\left[1,30\right]$, $p_{t}^{i},h_{t}^{i},b_{t}^{i}\in\left[1,10\right]$ and 
$k_{t}^{i}\in\left[1,50\right]$, the results even improve significantly as table \ref{t:tabla2} shows.

\begin{table}[htb]
\[
\begin{tabular}{|c|c|c|c|c|}
\hline
Players & $T=2$ & $T=3$ & $T=4$ & $T=5$ \\ \hline
2 & $99.984\%$ & $99.996\%$ & $99.999\%$ & $100\%$ \\ 
3 & $99.997\%$ & $99.995\%$ & $99.999\%$ & $99.999\%$ \\ 
4 & $99.998\%$ & $99.996\%$ & $99.999\%$ & $99.998\%$ \\ 
5 & $100\%$ & $99,999\%$ & $100\%$ & $100\%$ \\ 
6 & $100\%$ & $100\%$ & $100\%$ & $100\%$ \\ \hline
\end{tabular}%
\]
\caption{Percentage of instances fulfilling the condition of Theorem \ref{prop2 copy(1)} for instances with positive costs\label{t:tabla2}}
\end{table}

\renewcommand{\arraystretch}{}

In the previous simulation, the range of variation for the costs have been chosen so that those costs are actually  relevant to determine the optimal plans for each coalition. In addition, if the the set up costs are large compare to the other costs, as for instance for  $d_{t}^{i}\in\left[0,10\right]$, $p_{t}^{i},h_{t}^{i},b_{t}^{i}\in\left[0,10\right]$ and $k_{t}^{i}\in\left[50,500\right]$  the percentage of instances where the unitary Owen point is a core allocation is close to $99,995\%$, even for the case of two players and two periods. Moreover, if the demand is larger as it happens in the following situation  $d_{t}^{i}\in\left[10,50\right]$, 
$p_{t}^{i},h_{t}^{i},b_{t}^{i}\in\left[0,10\right]$ and $k_{t}^{i}\in\left[0,50\right]$, percentages of ``success'' also increase close to 1 ($99,999\%$). 
\medskip

The next example illustrates Proposition \ref{prop1} and Theorem \ref{prop2 copy(1)}. It shows how unitary Owen points are calculated by using unitary prices.

\begin{example}
Consider the following SI-situation with three periods and three players an
the associated SI-game: 
\begin{equation*}
\begin{array}{|c|c|c|c||c|c|c||c|c||c|c||c|c|c||c|}
\hline
& d_{1}^{_{S}} & d_{2}^{_{S}} & d_{3}^{_{S}} & p_{1}^{_{S}} & p_{2}^{_{S}} & 
p_{3}^{_{S}} & h_{1}^{_{S}} & h_{2}^{_{S}} & b_{1}^{_{S}} & b_{2}^{_{S}} & 
k_{1}^{_{S}} & k_{2}^{_{S}} & k_{3}^{_{S}} & c \\ \hline
\{1\} & 1 & 3 & 1 & 1 & 1 & 1 & 1 & 1 & 1 & 1 & 3 & 1 & 5 & 8 \\ \hline
\{2\} & 2 & 1 & 1 & 2 & 3 & 4 & 1 & 1 & 1 & 1 & 1 & 4 & 8 & 12 \\ \hline
\{3\} & 2 & 1 & 3 & 2 & 3 & 5 & 1 & 1 & 1 & 1 & 1 & 1 & 7 & 20 \\ \hline
\{1,2\} & 3 & 4 & 2 & 1 & 1 & 1 & 1 & 1 & 1 & 1 & 1 & 1 & 5 & 13 \\ \hline
\{1,3\} & 3 & 4 & 4 & 1 & 1 & 1 & 1 & 1 & 1 & 1 & 1 & 1 & 5 & 17 \\ \hline
\{2,3\} & 4 & 2 & 4 & 2 & 3 & 4 & 1 & 1 & 1 & 1 & 1 & 1 & 7 & 31 \\ \hline
\{1,2,3\} & 5 & 5 & 5 & 1 & 1 & 1 & 1 & 1 & 1 & 1 & 1 & 1 & 5 & 22 \\ \hline
\end{array}%
\end{equation*}

An optimal plan is the following: 
\begin{equation*}
\begin{array}{|c|c|c|c||c|c|c||c|}
\hline
& \sigma _{1}^{S} & \sigma _{2}^{S} & \sigma _{3}^{S} & P_{1}^{S}(\sigma
^{S}) & P_{2}^{S}(\sigma ^{S}) & P_{3}^{S}(\sigma ^{S}) & k(S) \\ \hline
\{1\} & 2 & 2 & 2 & 2 & 1 & 2 & 1 \\ \hline
\{2\} & 1 & 1 & 1 & 2 & 3 & 4 & 1 \\ \hline
\{3\} & 1 & 1 & 1 & 2 & 3 & 4 & 1 \\ \hline
\{1,2\} & 1 & 2 & 2 & 1 & 1 & 2 & 2 \\ \hline
\{1,3\} & 1 & 2 & 2 & 1 & 1 & 2 & 2 \\ \hline
\{2,3\} & 1 & 1 & 1 & 2 & 3 & 4 & 1 \\ \hline
\{1,2,3\} & 1 & 2 & 2 & 1 & 1 & 2 & 2 \\ \hline
\end{array}%
\end{equation*}

Thus, the unitary prices for the optimal plan above are:
\renewcommand{\arraystretch}{1.5} 
\begin{equation*}
\begin{array}{|c|c|c|c|}
\hline
& y_{1}\left( \sigma ^{S},d^{S},z^{S}\right) & y_{2}\left( \sigma
^{S},d^{S},z^{S}\right) & y_{3}\left( \sigma ^{S},d^{S},z^{S}\right) \\ 
\hline
\{1\} & 2+\frac{1}{5} & 1+\frac{1}{5} & 2+\frac{1}{5} \\ \hline
\{2\} & 2+\frac{1}{4} & 3+\frac{1}{4} & 4+\frac{1}{4} \\ \hline
\{3\} & 2+\frac{1}{6} & 3+\frac{1}{6} & 4+\frac{1}{6} \\ \hline
\{1,2\} & 1+\frac{1}{3} & 1+\frac{1}{6} & 2+\frac{1}{6} \\ \hline
\{1,3\} & 1+\frac{1}{3} & 1+\frac{1}{8} & 2+\frac{1}{8} \\ \hline
\{2,3\} & 2+\frac{1}{10} & 3+\frac{1}{10} & 4+\frac{1}{10} \\ \hline
\{1,2,3\} & 1+\frac{1}{5} & 1+\frac{1}{10} & 2+\frac{1}{10} \\ \hline
\end{array}%
\end{equation*}

One can observe that $y_{t}\left( \sigma ^{N},d^{N},z^{N}\right) \leq y_{t}\left( \sigma
^{S},d^{S},z^{S}\right) $ for all $t\in T$ and so, by Proposition \ref{prop1}, $\theta \left( \sigma ^{N},d^{N},z^{N}\right) =\left( 6.6,5.6,9.8\right) \in Core(N,c).$
\medskip

On the other hand, the ordering plan for the grand coalition $\widetilde{\sigma }^{N}=(1,2,3)$ belongs to 
an optimal plan and the associated unit prices are $y_{1}\left( \widetilde{\sigma }^{N},d^{N},z^{N}%
\right) =1+\frac{1}{5},y_{2}\left( \widetilde{\sigma }^{N},d^{N},z^{N}%
\right) =1+\frac{1}{5}$ and $y_{3}\left( \widetilde{\sigma }%
^{N},d^{N},z^{N}\right) =1+\frac{5}{5}.$ Note that for this plan $T^{N}=\{1,2,3\}.$
Theorem \ref{prop2 copy(1)} is here applied for the weights given in the
next table:
\renewcommand{\arraystretch}{1.5}
\begin{equation*}
\begin{array}{|c|c|c|c||c|}
\hline
& \beta _{1}^{_{S}}\geq & \beta _{2}^{_{S}}\geq & \beta _{3}^{_{S}}\geq & {%
\sum_{t\in T^{N}}\beta _{t}^{S}} \\ \hline
\{1\} & \frac{-4}{5} & \frac{3}{5} & 0 & \frac{-1}{5} \\ \hline
\{2\} & \frac{-8}{5} & \frac{-9}{5} & -2 & \frac{-27}{5} \\ \hline
\{3\} & \frac{-8}{5} & \frac{-9}{5} & -6 & \frac{-47}{5} \\ \hline
\{1,2\} & \frac{3}{10} & \frac{4}{10} & 0 & \frac{7}{10} \\ \hline
\{1,3\} & \frac{3}{10} & \frac{4}{10} & 0 & \frac{7}{10} \\ \hline
\{2,3\} & \frac{-16}{5} & \frac{-18}{5} & -8 & \frac{-27}{5} \\ \hline
\end{array}%
\end{equation*}%
Hence, it follows that $\theta \left( \widetilde{\sigma }^{N},d^{N},z^{N}\right) =\left(
6^{\prime }8,5^{\prime }6,9^{\prime }6\right) \in Core(N,c).$
\end{example}

This section is completed with a third example which shows that if any of the conditions either of the Proposition \ref{prop1} or Theorem \ref{prop2 copy(1)} fail, the unitary Owen points are not core allocations any more.

\begin{example}
\label{example-2P} Consider now the following SI-situation with three periods, two players, and the associated 2-player SI-game:

\begin{equation*}
\begin{array}{|c|c|c|c||c|c|c||c|c||c|c||c|c|c||c||}
\hline
& d_{1}^{_{S}} & d_{2}^{_{S}} & d_{3}^{_{S}} & p_{1}^{_{S}} & p_{2}^{_{S}} & 
p_{3}^{_{S}} & h_{1}^{_{S}} & h_{2}^{_{S}} & b_{1}^{_{S}} & b_{2}^{_{S}} & 
k_{1}^{_{S}} & k_{2}^{_{S}} & k_{3}^{_{S}} & c \\ \hline
\{1\} & 0 & 10 & 10 & 1 & 1 & 1 & 1 & 1 & 1 & 1 & 1 & 50 & 15 & 46 \\ \hline
\{2\} & 0 & 35 & 0 & 1 & 1 & 1 & 1 & 1 & 1 & 1 & 1 & 50 & 15 & 71 \\ \hline
\{1,2\} & 0 & 45 & 10 & 1 & 1 & 1 & 1 & 1 & 1 & 1 & 1 & 50 & 15 & 115 \\ 
\hline
\end{array}%
\end{equation*}
There is a single optimal ordering plan which is
\begin{equation*}
\begin{array}{|c|c|c|c||c|c|c||c|}
\hline
& \sigma _{1}^{S} & \sigma _{2}^{S} & \sigma _{3}^{S} & P_{1}^{S}(\sigma
^{S}) & P_{2}^{S}(\sigma ^{S}) & P_{3}^{S}(\sigma ^{S}) & k(S) \\ \hline
\{1\} & 0 & 1 & 3 & 0 & 2 & 1 & 16 \\ \hline
\{2\} & 0 & 1 & 0 & 0 & 2 & 0 & 1 \\ \hline
\{1,2\} & 0 & 2 & 2 & 0 & 1 & 2 & 50 \\ \hline
\end{array}%
\end{equation*}%
The unitary prices for the optimal plan above are: 
\renewcommand{\arraystretch}{1.5}
\begin{equation*}
\begin{array}{|c|c|c|c|}
\hline
& y_{1}\left( \sigma ^{S},d^{S}\right) & y_{2}\left( \sigma ^{S},d^{S}\right)
& y_{3}\left( \sigma ^{S},d^{S}\right) \\ \hline
\{1\} & 0 & 2+\frac{1}{10} & 1+\frac{15}{10} \\ \hline
\{2\} & 0 & 2+\frac{1}{35} & 0 \\ \hline
\{1,2\} & 0 & 1+\frac{50}{55} & 2+\frac{50}{55} \\ \hline
\end{array}%
\end{equation*}

Note that $\theta \left( \sigma ^{N},d^{N},z^{N}\right) =\left( 30+\frac{200}{11},35+%
\frac{350}{11}\right) =(48^{\prime }\widehat{18},66^{\prime }\widehat{81})$
is not a core allocation. Theorem \ref{prop2 copy(1)} fails here because $T^{N}=\{2\}$ and $\beta _{2}^{\{1\}}\geq \frac{5}{4}.$
\end{example}

\section{SI-games and PI-games}


To complete the paper we provide a relationship between a generic SI-game and a specific
family of PI-games through Owen's points of the latter. We use Owen
points from an \textit{ad hoc}  family of PI-situations constructed from core
allocations of the so called \textit{surplus game} which measures the excess in costs that
occurs with respect to the minimum unit price. This interesting relationship
simplifies the analysis and construction of core allocations for SI-games.

First, we introduce the minimum unitary prices for every optimal plan. Denote
by $\Delta :=\left( \sigma ^{S}\right)_{S\in \mathcal{P}(N)}$ an optimal
plan in $\Lambda (N,D,Z).$

\begin{definition}
Let $(N,D,Z)$ be a SI-situation. The minimum unitary price for $\Delta$, in
each period $t\in T$, is 
\begin{equation*}
y_{t}^{\ast }(\Delta )=\min_{\substack{ S\subseteq N  \\ d_{t}^{S}\neq 0}}%
\{y_{t}\left( \sigma ^{S},d^{S},z^{S}\right) \}.
\end{equation*}
\end{definition}

Second, for each coalition we measure the excess in costs that occurs with
respect to the minimum unit prices. The resulting cost game is what we have called
\emph{surplus game}.

\begin{definition}
Let $(N,D,Z)$ be a SI-situation and $(N,c)$ the associated SI-game. For any $%
\Delta \in \Lambda (N,D,Z)$, the \emph{surplus game} $(N,c^{\Delta })$ is
defined for all $S\subseteq N,$ as
\begin{equation*}
c^{\Delta}(S):=c(S)-\sum_{t=1}^{T}y_{t}^{\ast }(\Delta )\cdot d_{t}^{S}.
\end{equation*}
\end{definition}

Note that the surplus game is a non-negative cost game which measures the increase 
in costs by the influence of set-up costs. The first result of this section
shows that surplus games are always balanced.

\begin{proposition}
Every surplus game $(N,c^{\Delta })$ is balanced.
\end{proposition}

\begin{proof}
It follows from Theorem \ref{Balanced} that every SI-game $(N,c)$ is balanced.
Take a core allocation $x\in \mathbb{R}^{N}$ for it. For each $S\subset N$ it holds that 
\begin{eqnarray*}
x(S) &\leq &c(S)\Longleftrightarrow x(S)-\sum_{t=1}^{T}y_{t}^{\ast }(\Delta
)\cdot d_{t}^{S}\leq c(S)-\sum_{t=1}^{T}y_{t}^{\ast }(\Delta )\cdot d_{t}^{S}
\\
&\Longleftrightarrow &x(S)-\sum_{t=1}^{T}y_{t}^{\ast }(\Delta )\cdot
d_{t}^{S}\leq c^{\Delta }(S) \\
&\Longleftrightarrow &\sum_{i\in S}\left( x_{i}-\sum_{t=1}^{T}y_{t}^{\ast
}(\Delta )\cdot d_{t}^{i}\right) \leq c^{\Delta }(S).
\end{eqnarray*}

Moreover $x(N)=c(N)$ what easily implies that  $\sum_{i\in N}\left(
x_{i}-\sum_{t=1}^{T}y_{t}^{\ast }(\Delta )\cdot d_{t}^{i}\right) =c^{\Delta
}(N)$. Hence, we conclude that $\left( x_{i}-\sum_{t=1}^{T}y_{t}^{\ast }(\Delta )\cdot
d_{t}^{i}\right) _{i\in N}\in Core(N,c^{\Delta }).$
\end{proof}

In the following we use this game to construct core allocations for SI-games by means of the Owen points of the surplus game which is an easy PI-game. Next result provides 
a necessary and sufficient condition for this purpose: the set-up costs cannot contribute to any 
increase in costs for the grand coalition. In other words, there are no cost exceeding 
the unit prices of the grand coalition.

\begin{proposition}
Let $(N,c)$ be a SI-game. For any $\Delta \in \Lambda (N,D,Z),$
$$c^{\Delta }(N)=0 \mbox{ if and only if } \left( \sum_{t=1}^{T}y_{t}^{\ast }(\Delta
)\cdot d_{t}^{i}\right) _{i\in N}\in Core(N,c).$$
\end{proposition}

\begin{proof}
\textbf{(If)} If $c^{\Delta }(N)=0$ then $\sum_{t=1}^{T}y_{t}^{\ast }(\Delta )\cdot
d_{t}^{N}=c(N).$ For each $S\subset N,$ $\sum_{i\in S}\left(
\sum_{t=1}^{T}y_{t}^{\ast }(\Delta )\cdot d_{t}^{i}\right)
=\sum_{t=1}^{T}y_{t}^{\ast }(\Delta )\cdot d_{t}^{S}\leq
\sum_{t=1}^{T}y_{t}\left( \sigma ^{S},d^{S}\right) \cdot d_{t}^{S}=c(S).$
Thus, $\left( \sum_{t=1}^{T}y_{t}^{\ast }(\Delta )\cdot d_{t}^{i}\right)
_{i\in N}\in Core(N,c).$\newline
\textbf{(Only if)} If $\left( \sum_{t=1}^{T}y_{t}^{\ast }(\Delta )\cdot
d_{t}^{i}\right) _{i\in N}\in Core(N,c)$, it is satisfy that $%
\sum_{t=1}^{T}y_{t}^{\ast }(\Delta )\cdot d_{t}^{N}=c(N),$ and so $c^{\Delta
}(N)=0.$
\end{proof}
\medskip

The main theorem in this section shows that the core
of any SI-game consists of the Owen points of certain PI-games obtained from
core allocations of surplus games. To state this theorem, it is necessary to describe
a procedure to construct a PI-situation from core allocations of surplus games.
\medskip

Consider a SI-situation $(N,D,Z)$, the associated SI-game $(N,c)$, and 
the surplus game $(N,c^{\Delta })$, for $\Delta \in \Lambda (N,D,Z).$ For any $%
\alpha \in Core(N,c^{\Delta }),$ $\left( N,\overline{D}(\alpha ),\overline{Z}%
\right) $ is a PI-situation with $\overline{Z}=(\overline{K},\overline{H},%
\overline{B},\overline{P})$ and $\overline{D}(\alpha )=[\overline{d}%
^{1},\ldots ,\overline{d}^{n}]^{\prime },\overline{K}=0,\overline{H}%
=[M,\ldots ,M]^{\prime },\overline{B}=[M,\ldots ,M]^{\prime },\overline{P}=[%
\overline{p},\ldots ,\overline{p}]^{\prime },$ with $\overline{p}%
=(y_{1}^{\ast }(\Delta ),...,y_{T}^{\ast }(\Delta ),1),$ $\overline{d}%
^{i}=(d_{1}^{i},...,d_{T}^{i},\alpha _{i})$ for all $i\in N$ and $M\in 
\mathbb{R}^{N}$ large enough. This procedure shows that any SI-situation
can be transformed into multiple PI-situations just by using the core of the
surplus games.

\begin{theorem}
Let $(N,c)$ be a SI-game and $(N,c^{\Delta })$ the associated surplus game
for $\Delta \in \Lambda (N,D,Z).$ Thus, 
\begin{equation*}
Core(N,c)=\left\{ Owen\left( N,\overline{D}(\alpha ),\overline{Z}%
\right): \alpha \in Core(N,c^{\Delta })\right\} .
\end{equation*}
\end{theorem}

\begin{proof}
As $(N,c^{\Delta })$ is balanced, there is at least one $\alpha \in \mathbb{R}%
^{N}$, such that $\alpha (S)\leq c^{\Delta
}(S)=c(S)-\sum_{t=1}^{T}y_{t}^{\ast }(\Delta )\cdot d_{t}^{S}$ for all $%
S\subset N$ and $\alpha (N)=c(N)-\sum_{t=1}^{T}y_{t}^{\ast }(\Delta )\cdot
d_{t}^{N}.$ Consider a PI-situation $(N,\overline{D}(\alpha ),\overline{Z})$
with $T+1$ periods, where%
\begin{equation*}
\overline{D}(\alpha )=[\overline{d}^{1},\ldots ,\overline{d}^{n}]^{\prime },%
\overline{K}=0,\overline{H}=[M,\ldots ,M]^{\prime },\overline{B}=[M,\ldots
,M]^{\prime },\overline{P}=[\overline{p},\ldots ,\overline{p}]^{\prime }
\end{equation*}
with $\overline{p}=(y_{1}^{\ast }(\Delta ),...,y_{T}^{\ast }(\Delta ),1),$ $%
\overline{d}^{i}=(d_{1}^{i},...,d_{T}^{i},\alpha _{i})$ for all $i\in N$ and 
$M\in \mathbb{R}^{N}$ large enough. For each $i\in N,$ $Owen_{i}\left( N,%
\overline{D}(\alpha ),\overline{Z}\right) =\sum_{t=1}^{T+1}y_{t}^{\ast
}(N)d_{t}^{i}=\left( \sum_{t=1}^{T}y_{t}^{\ast }(\Delta )d_{t}^{i}\right)
+\alpha _{i}.$ Then, for all $S\subset N$: 
\begin{eqnarray*}
\sum\limits_{i\in S}Owen_{i}\left( N,\overline{D}(\alpha ),\overline{Z}%
\right) &=&\sum_{t=1}^{T}y_{t}^{\ast }(\Delta )\cdot d_{t}^{S}+\alpha (S) \\
&\leq &\sum_{t=1}^{T}y_{t}^{\ast }(\Delta )\cdot
d_{t}^{S}+c(S)-\sum_{t=1}^{T}y_{t}^{\ast }(\Delta )\cdot d_{t}^{S}=c(S).
\end{eqnarray*}

Moreover, $\sum\limits_{i\in N}Owen_{i}\left( N,\overline{D}(\alpha ),%
\overline{Z}\right) =\sum_{t=1}^{T}y_{t}^{\ast }(\Delta )d_{t}^{N}+\alpha
(N)=\sum_{t=1}^{T}y_{t}^{\ast }(\Delta
)d_{t}^{N}+c(N)-\sum_{t=1}^{T}y_{t}^{\ast }(\Delta )d_{t}^{N}=c(N).$ Thus
$Owen\left( N,\overline{D}(\alpha ),\overline{Z}\right) \in Core(N,c).$
\smallskip

On the other hand, if $x\in Core(N,c),$ for each $S\subset N$ it holds  
\begin{eqnarray*}
x(S) &\leq &c(S); \\
x(S)-\sum_{t=1}^{T}y_{t}^{\ast }(\Delta )\cdot d_{t}^{S} &\leq
&c(S)-\sum_{t=1}^{T}y_{t}^{\ast }(\Delta )\cdot d_{t}^{S}; \\
\sum\limits_{i\in S}x_{i}-\sum\limits_{i\in S}\left(
\sum_{t=1}^{T}y_{t}^{\ast }(\Delta )\cdot d_{t}^{\{i\}}\right) &\leq
&c^{\Delta }(S); \\
\sum\limits_{i\in S}\left( x_{i}-\sum_{t=1}^{T}y_{t}^{\ast }(\Delta )\cdot
d_{t}^{\{i\}}\right) &\leq &c^{\Delta }(S);
\end{eqnarray*}

Moreover $x(N)=c(N)\Leftrightarrow \sum\limits_{i\in N}\left(
x_{i}-\sum_{t=1}^{T}y_{t}^{\ast }(\Delta )\cdot d_{t}^{i}\right) =c^{\Delta
}(N).$ Thus for each $x\in Core(N,c)$ we can take $\alpha
_{i}:=x_{i}-\sum_{t=1}^{T}y_{t}^{\ast }(\Delta )\cdot d_{t}^{i}$ for all $%
i\in N$ such that $\alpha \in Core(N,c^{\Delta }).$  From here it easily follows that $Owen\left( N,%
\overline{D}(\alpha ),\overline{Z}\right) =\left( \left(
\sum_{t=1}^{T}y_{t}^{\ast }(\Delta )\cdot d_{t}^{i}\right) +\alpha
_{i}\right) _{i\in N}=x.$
\end{proof}
\medskip 

We illustrate the procedure above  with the Example \ref{example-2P} shown above.

\begin{example} Consider the 2-player SI-game given in Example \ref{example-2P}.
We have shown that the unitary Owen point is not a core allocation for this example. 
It can be easily checked that the minimal unit prices are: 
\renewcommand{\arraystretch}{1.5}
\begin{equation*}
\begin{array}{|c|c|c|}
\hline
y_{1}^{\ast }(\Delta ) & y_{2}^{\ast }(\Delta ) & y_{3}^{\ast }(\Delta ) \\ 
\hline
0 & 1+\frac{50}{55} & 1+\frac{15}{10} \\ \hline
\end{array}%
\end{equation*}

Thus, the surplus game is given by
\renewcommand{\arraystretch}{1.5}
\begin{equation*}
\begin{array}{|c|c|c|}
\hline
& c & c^{\Delta } \\ \hline
\{1\} & 46 & 1+\frac{10}{11} \\ \hline
\{2\} & 71 & 4+\frac{2}{11} \\ \hline
\{1,2\} & 115 & 4+\frac{1}{11} \\ \hline
\end{array}%
\end{equation*}

Consider a core allocation from the surplus game, for instance the
nucleolus $\eta (N,c^{\Delta })=(\frac{10}{11},3\frac{2}{11}).$ We  
obtain a core allocation for the SI-game just by calculating the Owen point
of the associated PI-situation $\left( N,\overline{D}(\eta (N,c^{\Delta })), 
\overline{Z}\right).$ Thus, $Owen\left( N,\overline{D}(\eta (N,c^{\Delta
})),\overline{Z}\right) =(45,70).$ One can conclude that  
\begin{equation*}
\eta (N,c)=Owen\left( N,\overline{D}(\eta (N,c^{\Delta })),\overline{Z}%
\right).
\end{equation*}
\end{example}

In the above example, the nucleolus of the surplus game leads to the
nucleolus of the SI-game through the Owen point. The last result in the
paper shows that this close relationship between both nucleoli always holds, 
e.g. the nucleolus of any SI-game matches the Owen point for the
PI-situation obtained from the nucleolus of the surplus game.

\begin{proposition}
Let $(N,D,Z)$ be a SI-situation, $(N,c)$ the associated SI-game, and $%
(N,c^{\Delta })$ the surplus game for $\Delta \in \Lambda (N,D,Z).$ Thus, 
\begin{equation*}
Owen\left( N,\overline{D}(\eta (N,c^{\Delta })),\overline{Z}\right) =\eta
(N,c).
\end{equation*}
\end{proposition}

\begin{proof}
It is known that $x\in Core(N,c)$ if and only if $x^{\Delta }:=\left(
x_{i}-\sum_{t=1}^{T}y_{t}^{\ast }(\Delta )\cdot d_{t}^{i}\right) _{i\in
N}\in Core(N,c^{\Delta })$. Thus the excess vectors, $e(S,x)$, and $e_{\Delta
}(S,x^{\Delta })$ coincide.

For each coalition $S\subseteq N$, it holds that:

\begin{eqnarray*}
e(S,\eta (N,c)) &=&c(S)-\sum\limits_{i\in S}\eta _{i}(N,c) \\
&=&c^{\Delta }(S)+\sum_{t=1}^{T}y_{t}^{\ast }(\Delta )\cdot
d_{t}^{S}-\sum\limits_{i\in S}\eta _{i}(N,c) \\
&=&c^{\Delta }(S)-\sum\limits_{i\in S}\left( \eta
_{i}(N,c)-\sum_{t=1}^{T}y_{t}^{\ast }(\Delta )\cdot d_{t}^{i}\right).
\end{eqnarray*}

Therefore, $\eta _{i}(N,c^{\Delta })=\eta
_{i}(N,c)-\sum_{t=1}^{T}y_{t}^{\ast }(\Delta )\cdot d_{t}^{i}$ \ for all $i\in N$ 
because otherwise $\eta (N,c)$ would not be the nucleolus.
Moreover, 
\begin{eqnarray*}
c^{\Delta }(S)-\eta _{i}(N,c^{\Delta }) &=&c(S)-\left(
\sum_{t=1}^{T}y_{t}^{\ast }(\Delta )\cdot d_{t}^{S}+\sum\limits_{i\in S}\eta
_{i}(N,c^{\Delta })\right) \\
&=&c(S)-\sum\limits_{i\in S}Owen_{i}\left( N,\overline{D}(\eta (N,c^{\Delta
})),\overline{Z}\right) \\
&=&e(S,Owen\left( N,\overline{D}(\eta (N,c^{\Delta })),\overline{Z}\right) ).
\end{eqnarray*}

This implies that $Owen\left( N,\overline{D}(\eta (N,c^{\Delta })),\overline{Z}\right)
=\eta (N,c).$
\end{proof}

\section{Concluding Remarks}

The study of cooperation in lot-sizing problems with backlogging and
heterogeneous costs begins in Guardiola et al. (2020). The authors prove
that there are always fair allocations of the overall operation cost among
the firms so that no group of agents profit from leaving the consortium.
They propose a parametric family of cost allocations and provide
sufficient conditions for this to be a stable family against coalitional
defections of firms and focus on those periods of the time horizon
that are consolidated analyzying their effect on the stability of cost
allocations.

To complete the study of those cooperative lot-sizing models, this paper 
presents the unitary Owen point. As mentioned, the Owen point works great 
for constructing core-allocations in the class of PI-games. Unfortunately, this does not work for SI-problems any more. In spite of that, we manage here to construct a particular kind of prices, that we call unitary prices, based on the
production, inventory and backlogging costs and a proportion of the fixed
order cost which depends on the total demand satisfied in each period. 
These unit prices enable one to replicate the construction of the Owen point
so that these allocations ``a la Owen" are called unitary Owen points.
We provide necessary and sufficient conditions for unitary
Owen points to be core-allocations. Moreover, we show a relationship
between SI-games and a certain family of PI-games through Owen's points of
the latter. Specifically, we use Owen points from a family of PI-situations, constructed 
from core allocations of an \textit{ad hoc} surplus game  which
measures the excess in costs that occurs with respect to the minimum unit
prices. This amazing relationship enables one to easily construct another family of fair
allocations for SI-games. 

\section*{Acknowledgments}

The research of the second author is supported from Spain's 
Ministerio de Ciencia, Innovaci\'{o}n y Universidades (MCIU), from the
Agencia Estatal de Investigaci\'{o}n (AEI) and from the Fondo Europeo de
Desarrollo Regional (FEDER) under the project {PGC2018-097965-B-I00}. The
research of the third author is partially supported from Spanish
Ministry of Education and Science/FEDER grant number {MTM2016-74983-C02-01},
and from  projects {FEDER-US-1256951}, {CEI-3-FQM331} and \textit{NetmeetData}:
Ayudas Fundaci\'{o}n BBVA a equipos de investigaci\'{o}n cient\'{\i}fica
2019.

\end{document}